\documentclass{ws-rv9x6}
\usepackage{ws-rv-van}             

\makeindex
\begin{document}
\setcounter{chapter}{2}
\chapter{Darwinian purifying selection versus complementing strategy in Monte Carlo simulations}

\author[W. Waga, M. Zawierta, J.Kowalski, S. Cebrat]{Wojciech Waga${}^1$, Marta Zawierta${}^1$, Jakub Kowalski${}^1$, Stanis{\l}aw Cebrat${}^1$}

\address{${}^1$ Department of Genomics, Faculty of Biotechnology, University of Wroc{\l}aw, ul. Przybyszewskiego 63/77, 51-148 Wroc{\l}aw, Poland. 
E-mail: cebrat@smorfland.uni.wroc.pl.}

\begin{abstract}
Intragenomic recombination (crossover) is a very important evolutionary mechanism. The crossover events are not evenly distributed along the natural chromosomes. Monte Carlo simulations revealed that frequency of recombinations decides about the strategy of chromosomes' and genomes' evolution. In large panmictic populations, under high recombination rate the Darwinian purifying selection operates keeping the fraction of defective genes at the relatively low level. In small populations and under low recombination rate the strategy of complementing haplotypes seems to be more advantageous. Switching between the two strategies has a character of phase transition - it depends on inbreeding coefficient and crossover rate. The critical recombination rate depends also on the size of chromosome. It is also possible, that in one genome some chromosomes could be under complementing while some other under purifying selection. Such situation stabilizes genome evolution and reproduction strategy. It seems that this phenomenon can be responsible for the positive correlation between kinship and fecundity, recently found in the Islander population. When large population is forced to enter the complementing strategy, the phenomenon of sympatric speciation is observed.
\end{abstract}

\body

\section{Introduction}
It is difficult to define the term ``species'' in a way that applies to all naturally occurring populations. Usually a species is defined as ``all individual organisms of a natural population which interbreed at maturity in the wild and whose interbreeding produces fertile offspring''. Note words: natural populations, the wild and fertile offspring. In unnatural conditions, a lot of barriers can be broken - in captivity or in laboratories, interbreeding of individuals belonging to different species could be successful. On the other hand, mule is not a species, because mule mating with mule never gives an offspring. Another definition of species is ``a pool of genes which could be freely exchanged between individuals belonging to the population''. The last definition is particularly plausible for geneticists. 
Species could be extinct thus, species have to appear. Appearing of new species is called speciation. There are two distinct ways of speciation: allopatric and sympatric speciation. There are no problems with allopatric speciation when populations of one species are divided by physical, geographical or biological barriers, and they even physically cannot interbreed for a long time. They eventually form two (or more) different species which cannot produce fertile offspring by interbreeding - such a way of emerging of new species is called allopatric speciation. 

Sympatric speciation is a quite different phenomenon - it is an emergence of a new species inside the older one on the same territory without any physical or geographical barriers. Since Ernst Mayr \cite{Mayr}, who was rather sceptical about the possibility of sympatric speciation, this phenomenon is still debatable \cite{Jiggins}, though some well documented empirical data that supports it already exists \cite{Barluenga},\cite{Bunje}. Also some theoretical models that address the problems showed the possibility of speciation in sympatry \cite{Doebeli}, \cite{Stauffer}.
In this section we are going to show in computer modelling that sympatric speciation could be the intrinsic property of the evolving populations.

\section{Mutations, frequency of defective alleles in the populations and complementation}

Mutation is a change in a sequence of nucleotides in a genome, it can transform the functional gene into its defective unfunctional form see \cite{Dorota}.  Many experimental and theoretical analyses suggest that the level of mutational pressure is of the order of one mutation per genome per generation, independently of the genome size \cite{Azbel}, \cite{Drake}. It is even possible, that there is a phase transition between the ``order phase'' where the mutational pressure is low enough to enable the population survival and ``disorder phase'' where the mutations happen too frequently to be effectively eliminated by selection \cite{Paulo}. According to neo-Darwinism, mutations are random thus, they could happen in any gene and it is selection which eliminates some of them with higher efficiency than others. If homozygous state (when individual has both defective alleles in the locus) is lethal then, in the Mendelian populations (infinite in size), where we assume random mutations, random crossovers with high frequency and no mating preferences we should expect for lethal mutations:
\begin{equation}
p_d=f^2, 
\end{equation}
where:
\begin{itemize}
\item[$p_d$] - probability of genetic death caused by alleles of single locus,
\item[$f$] - fraction of defective alleles in this locus in the whole genetic pool of interbreeding population.
\end{itemize}
Thus, probability of death corresponds to probability of meeting two defective alleles at the same position in one diploid genome. It depends on the frequency of defects at these positions in all bitstrings (haplotypes) in the population (genetic pool). Since selection eliminates defective alleles from the genetic pool by the genetic death, their frequency would decrease with time if there are no new mutations. Nevertheless, mutational pressure introduces new defective alleles into the genetic pool and, after long time of evolution equilibrium should be established with a level of defective genes corresponding to the force of selection and mutational pressures. The phenomenon of elimination the defective genes from the genetic pool of population is called the purifying selection. One can expect that if all genes have the same selection value and the same probability of mutations, the frequency of defects in the genetic pool should be at the same, rather low level, in all loci. 

In the situation described in the above paragraph the effect of defective gene can be invisible - phenotype of the organism is normal if only one copy of the two corresponding alleles is defective and the defect is recessive. It is said that the mutation or defective gene is complemented by the normal, functional one. In our model we assume that all defective genes are complemented by their normal counterparts. That is why it is imaginable at the extreme situation that at each locus of the diploid genome composed of two bitstrings one bit is set to 0 while the other one is set to 1 (see fig. ). Such a structure of genomes is called the complementing haplotypes. Of course it is highly improbable in the model described above for the Mendelian populations. We have described in the previous chapter, that nonrandom distribution of genes on chromosomes generates some preferences in elimination of defective genes \cite{Biecek3}. In the Penna model \cite{Penna}, the differences in the effectiveness of elimination the defective alleles depend on the period of life when they are expressed. 

What  can we find in the natural populations?  
There are many known examples of very high frequencies of recessive defective genes in the human population i.e. cystic fibrosis, sickle cell anaemia or Tay-Sachs disorder. In the Caucasian race, one in every 25 persons is a carrier of a defect responsible for cystic fibrosis, among African-Americans one in 12 is a carrier of S haemoglobin gene responsible for sickle cell anaemia and one in every 27 of Ashkenazi American Jews is a carrier of defective Hex-A gene responsible for Tay-Sachs disorder (carrier is an individual with one defective, recessive allele and a second one in the corresponding locus - correct). All these examples have one common property - the incidence of the defective alleles is not universal in the human populations, they have rather endemic character or are characteristic for some ethnic groups. What are the reasons of such a high frequency and such an uneven distribution of these defects? 
The best characterized is a sickle cell anaemia (see \cite{Kwiatkowski} for review). The point mutation in a gene coding for haemoglobin results in a replacement of one, proper amino acid by a wrong one and the ``wrong'' haemoglobin is called S haemoglobin.  See the chapter 1 for an explanation on how single mutations in DNA cause the amino acid substitutions in proteins \cite{Dorota}. People possessing one gene for S haemoglobin and the other one for normal haemoglobin are carriers and it is said they have sickle cell trait. These people are normal in almost all respects and rarely develop problems related to their genetic conditions thus, we can state that the mutation is recessive. Could we say that selection doesn't discriminate the carriers? 
Sickle cell trait provides a survival advantage over people with normal haemoglobin in regions where malaria is endemic. In fact this trait provides neither absolute protection nor resistance to the disease. People, and particularly children, infected with \textit{Plasmodium falciparum} (which is a parasite responsible for malaria) are more likely to survive the acute illness if they have S haemoglobin gene - have a sickle cell trait. Each year, malaria attacks about 400 million people, two to three million of whom succumb to the illness. From our point of view, each year, natural selection (malaria) tests 400 million people discriminating the normal ones and favouring the carriers of S haemoglobin genes.

The other example is cystic fibrosis - carriers suppose to be more resistant to the infections of the alimentary tract. In a case of Tay Sachs disease interpretation is not so simple, some people suggest that carriers are more intelligent \cite{Cochran}, \cite{Spyropoulos}. Even if we assume it is true - does intelligence increase the survival probability and fertility? 

Taking into account the above examples it is possible to model the evolution of population where selection favours heterozygotes (carriers) which have one correct allele and one defective while defective homozygotes (both alleles in the corresponding loci are defective) are lethal.

\section{Assumptions of the model}
Our virtual population is composed of N individuals. Each individual is represented by their diploid genome - two bitstrings L bits long (haplotypes). Bits at the same positions (loci) in both bitstrings  correspond to alleles. Bit set to 0 corresponds to the wild allele (correct one) while bit set to 1 corresponds to the defective allele. All defective alleles are recessive, which means that both alleles at the corresponding positions should be defective to determine the defective phenotypic character. Each phenotypic defect is lethal - an individual with such a trait has to die. The whole genomes of newborns are checked at birth. If at any position of bitstrings both alleles are defective, the individual is eliminated otherwise it stays at the population. One can notice that in this model there are no ageing and all positions in the bitstrings seems to be equivalent. After each Monte Carlo step (MCs) the declared fraction of a population is randomly killed (usually $5\%$). To fill up the gap, in the next MCs randomly chosen individuals can reproduce. To produce an offspring, both chosen individuals produce the gamete: they copy their two haplotypes, introduce mutations into each copy of haplotype with a probability M and performs a crossover between the pair of new haplotypes with a probability C. Mutation replaces 0 bit by 1, if the bit chosen for mutation is already 1 it stays 1 - there are no reversions. Two gametes, each produced by one of the two parental individuals are joined and form a genome of the newborn which is immediately checked for its genetic status. The MCs is completed when the population reaches the declared size of N individuals. The main difference between this model and the Penna model is in the way how the genome is checked - in the Penna model it is checked chronologically, in this model all loci are checked at birth. 

\section{Positive selection for heterozygosity}
To show how the frequency of defective alleles could be affected by positive selection for heterozygous loci an additional condition has been introduced: the chance for reproduction of the organism is proportional to (h+1) where h represents the number of heterozygous loci in the genome (in a heterozygous loci alleles have different values - one allele is set to 0 and the second one is set to 1). Simulations have been performed for two different strategies of reproduction - cloning, asexual reproduction without any exchange of information between parents and, sexual reproduction as described in the above section but without recombination between haplotypes during the gamete production - (C=0). Notice that there are two possible modes of sexual reproduction for C=0. In the first one, bitstrings are enumerated - the first one and the second one. The first parental genome is a donor of a copy of its first bitstring and the second parent is a donor of a copy of its second bitstring. In this situation there are two separate sets of bitstrings. In the second mode parents are donors of randomly chosen copies of their bitstrings.

\begin{figure}
\centerline{\psfig{file=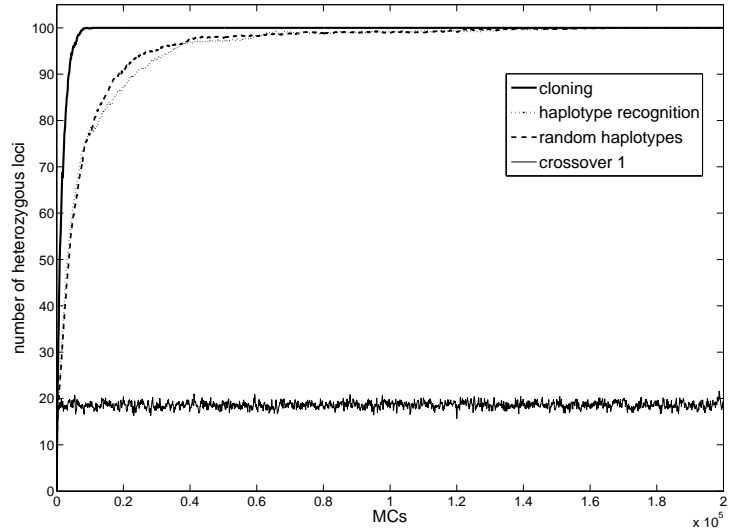,width=1\textwidth}}
  \caption{Evolution of genomes under different strategies of reproduction with a positive selection for heterozygous loci. Recombination between haplotypes prevents accumulation of defective genes. Cloning or production of gametes without recombination leads to the high level (100\%) of heterozygous loci.}
  \label{Figure:fig1}
\end{figure}

Simulations were performed for parameters: L=100, N=1000, M=1. The results of simulations are shown in Fig.\ref{Figure:fig1}. In a case of cloning, after relatively short time of simulation, all genomes became heterozygous in all loci. The results seem to be trivial because all homozygous defective loci eliminate the genomes (individuals) from the population while heterozygous loci increase the probability of their reproduction. The results of the sexual reproduction without crossover are not so obvious. Nevertheless, after longer simulations (200 000 MCs) all genomes become heterozygous in all loci which means that the frequency of defective alleles in the genetic pool is 0.5 for all loci. If we assume that the mutations are distributed randomly along the haplotypes in the whole genetic pool, then the probability of forming the killing configuration in a single locus ($1\|1$) is 0.25 while surviving - 0.75. Thus, probability of forming the surviving offspring should be of the order of $0.75^{100}\approx10^{-13}$. Populations would have no chance to survive under such conditions. Nevertheless, our virtual populations are not going to die. We have checked the Hamming distance (H) between all haplotypes in the populations. Hamming distance is a sum of differences between bits at the same position of bitstring counted for the whole length of the compared pair of bitstrings - it corresponds to the number of heterozygous loci in the genome in our case. After long simulations the Hamming distance for any combination of haplotypes equals 100 or 0 which means that in the whole populations only two types of haplotypes occur. Let us call them $T1$ and $T2$. $T1$ haplotype perfectly complements the $T2$ one thus, $T1||T2$ and $T2||T1$ survive. But $T1\||T1$ and $T2\||T2$ combinations are lethal. Nevertheless, the probability of forming the surviving configuration of haplotypes in such a population is $0.5$ instead of $10^{-13}$ in case of random distribution of defects on haplotypes. That is why population evolves in the direction of complementing the haplotypes rather than in the direction of purifying selection. In the first mode of the sexual reproduction we have two sets of haplotypes because we declared that the first parent introduces into the gamete the copy of its first haplotype and the second parent - the copy of its second haplotype. Thus, we have introduced some kind of haplotype recognition. Is it responsible for choosing the complementation strategy by evolving populations? 
To check this possibility we have not declared such recognition in the second mode of sexual version. In that strategy, the gametes were produced of randomly drawn parental haplotypes. The results resembled the previous ones - populations evolved toward the complementing haplotypes, though the dynamics of this evolution is different than in case of haplotype recognition, which will be described later in details. 
In the further series of simulations the recombination between haplotypes (C=1) was introduced into the sexual reproduction. In this case the populations choose the purifying selection and the level of defective alleles in the haplotypes was kept low. Thus, even if selection favors the heterozygosity - when recombination is high the populations choose the purifying selection. 
One can argue that under the reproduction without recombination between haplotypes the complementation is driven by selection favoring the heterozygosity. 

\section{Is the complementation strategy possible without an advantage of heterozygosity?}

\begin{figure}
\centerline{\psfig{file=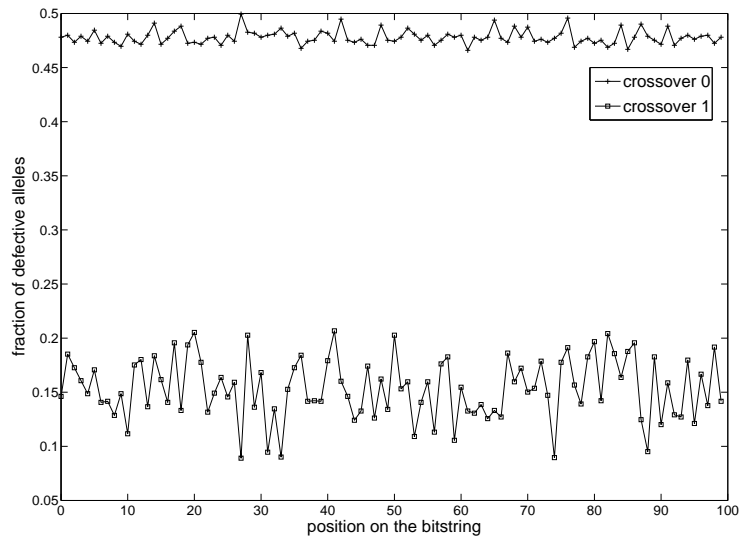,width=1\textwidth}}
  \caption{Distribution of defective genes along the (bitstrings) haplotypes in populations evolving without recombination or with 1 recombination per gamete production.}
  \label{Figure:fig2}
\end{figure}

The results presented at the end of the last section of the chapter describing the Penna model \cite{Biecek3} suggest that complementation strategy is possible even if there is no declared advantage of heterozygous loci. To show this phenomenon in the simplified model without age structure, we have repeated the simulations described in the above section but without declaring any preference for heterozygosity. All simulations with cloning or without recombination during the gamete production gave the populations with complementing haplotypes. The only simulation showing the Darwinian purifying selection was that one with recombination between haplotypes before production of gametes. The distribution of defective alleles along the haplotypes after the evolution with and without the crossover is shown in Fig.\ref{Figure:fig2}.
The results clearly suggest that there are two possible strategies of genomes' evolution. The first one based on Darwinian selection eliminating defective phenotypes and the second one which allows higher accumulation of defective alleles but it requires the complementation of defective alleles by the wild forms of the genes. The second strategy leads to appearing of specific and unique distribution of defective genes along the haplotypes. Thus, two problems arise, now:

\begin{itemize}
\item the first one is connected with that ``uniqueness'' - each simulation produces specific sets of complementary haplotypes - specific means that distribution of defective alleles along the haplotypes is unique and specific for each independent simulation. The probability that haplotypes from independently simulated populations would complement is equal to $0.5^{L}$, where L is the number of loci in the haplotype. Geneticists would say that crossbreeding between two such populations gives no surviving offspring or unfertile offspring - these two populations suppose to be different species, 
\item the second problem arises with the efficiency of recombination. If the crossover frequency of the order of 1 per one pair of haplotypes is enough to push the population toward the purifying selection and crossover 0 favours the complementing strategy - then, how the transition between these two strategies looks like?
\end{itemize}

\section{Phase transition between purifying selection and complementation strategy}

\begin{figure}
\centerline{\psfig{file=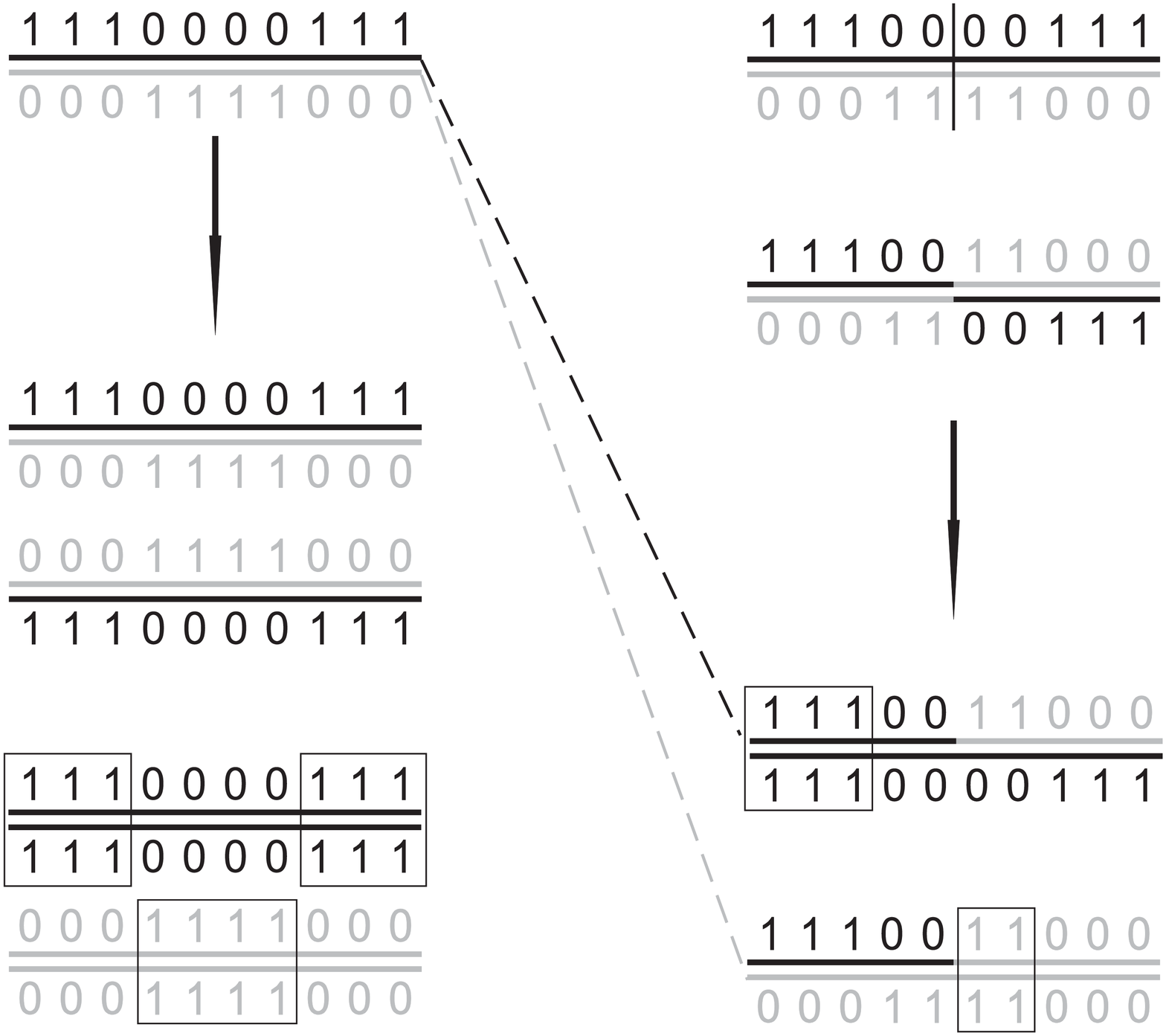,width=0.6\textwidth}}
  \caption{Why recombination prevents the complementation strategy?. If in the population evolving without recombination (left part) only two types of haplotypes occur (bold and light) the probability of forming the surviving newborn equals 0.5. But if in one individual during the gamete production recombination happens, then the possibility of forming a surviving newborn is negligible. The gamete has to find a proper gamete after recombination at the same point.}
  \label{Figure:fig3}
\end{figure}

\begin{figure}
\centerline{\psfig{file=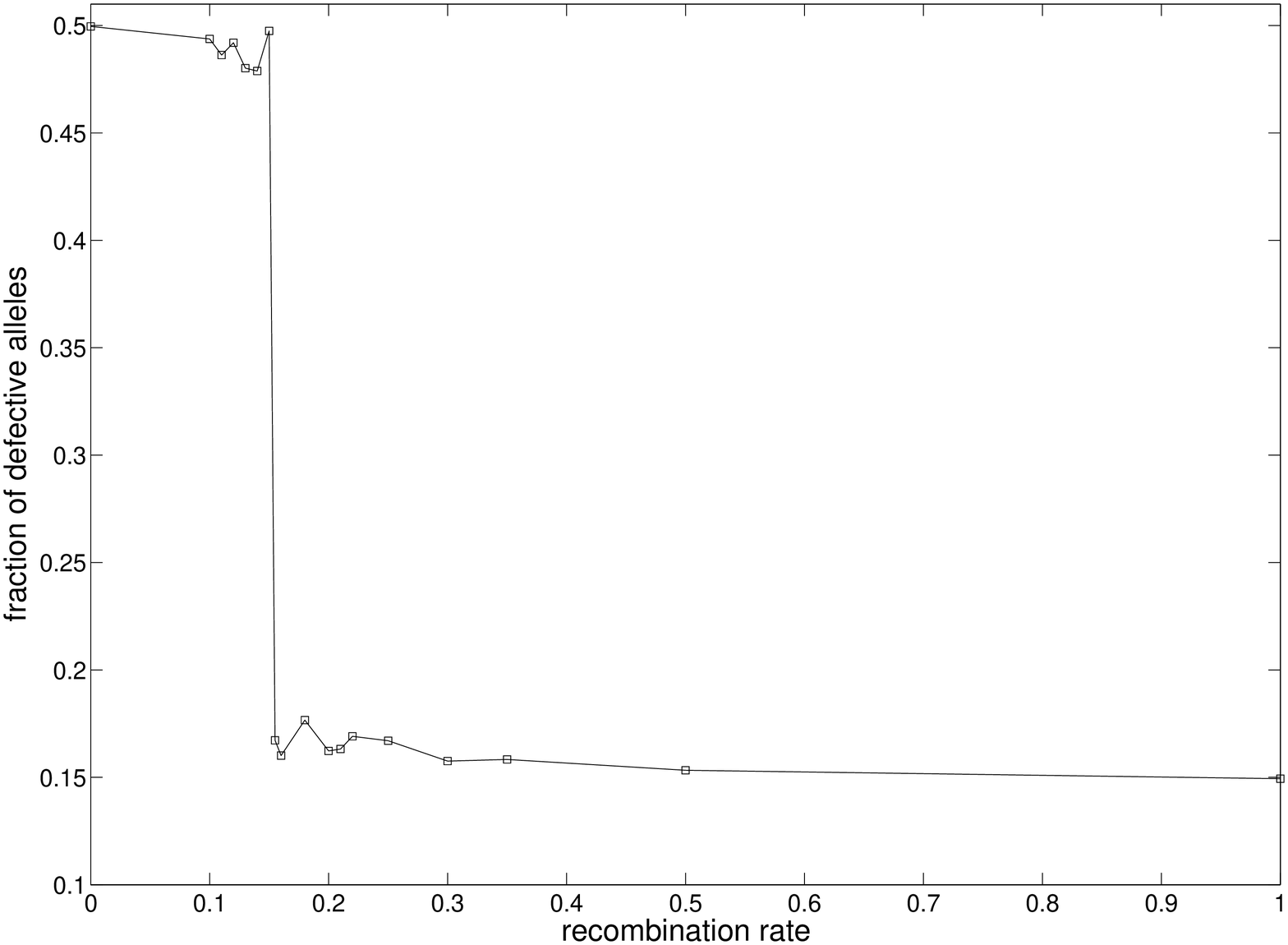,width=1\textwidth}}
  \caption{Effect of different recombination rate on the frequency of defective alleles (and the strategy of genome evolution). Parameters of simulations: L=100, M=1, N=1000.}
  \label{Figure:fig4}
\end{figure}

\begin{figure}
\centerline{\psfig{file=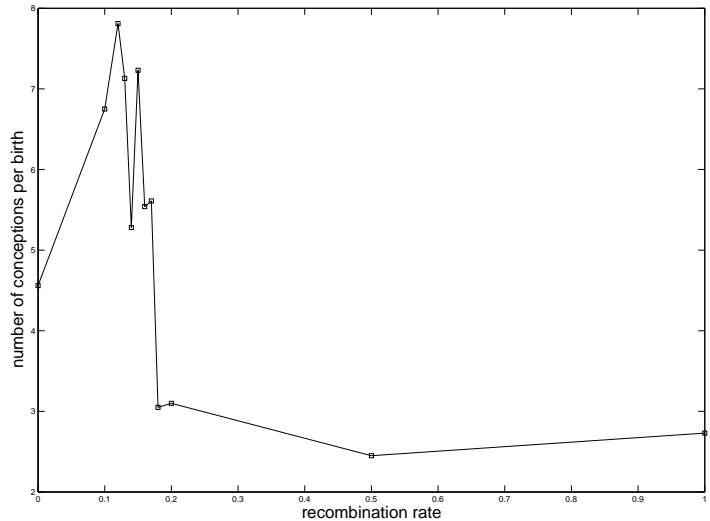,width=1\textwidth}}
  \caption{The relation between the recombination rate and average number of conceptions per one successful birth.}
  \label{Figure:fig5}
\end{figure}

\begin{figure}
\centerline{\psfig{file=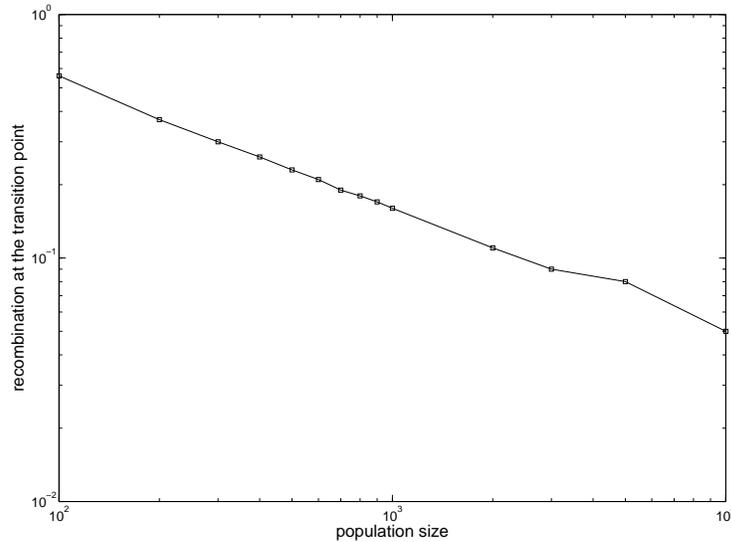,width=1\textwidth}}
  \caption{The relation between the frequency of recombination at the transition point (see Fig.\ref{Figure:fig4} and Fig.\ref{Figure:fig5}) and the populations' sizes. Notice the logarithmic scales of both axes.}
  \label{Figure:fig6}
\end{figure}

\begin{figure}
\centerline{\psfig{file=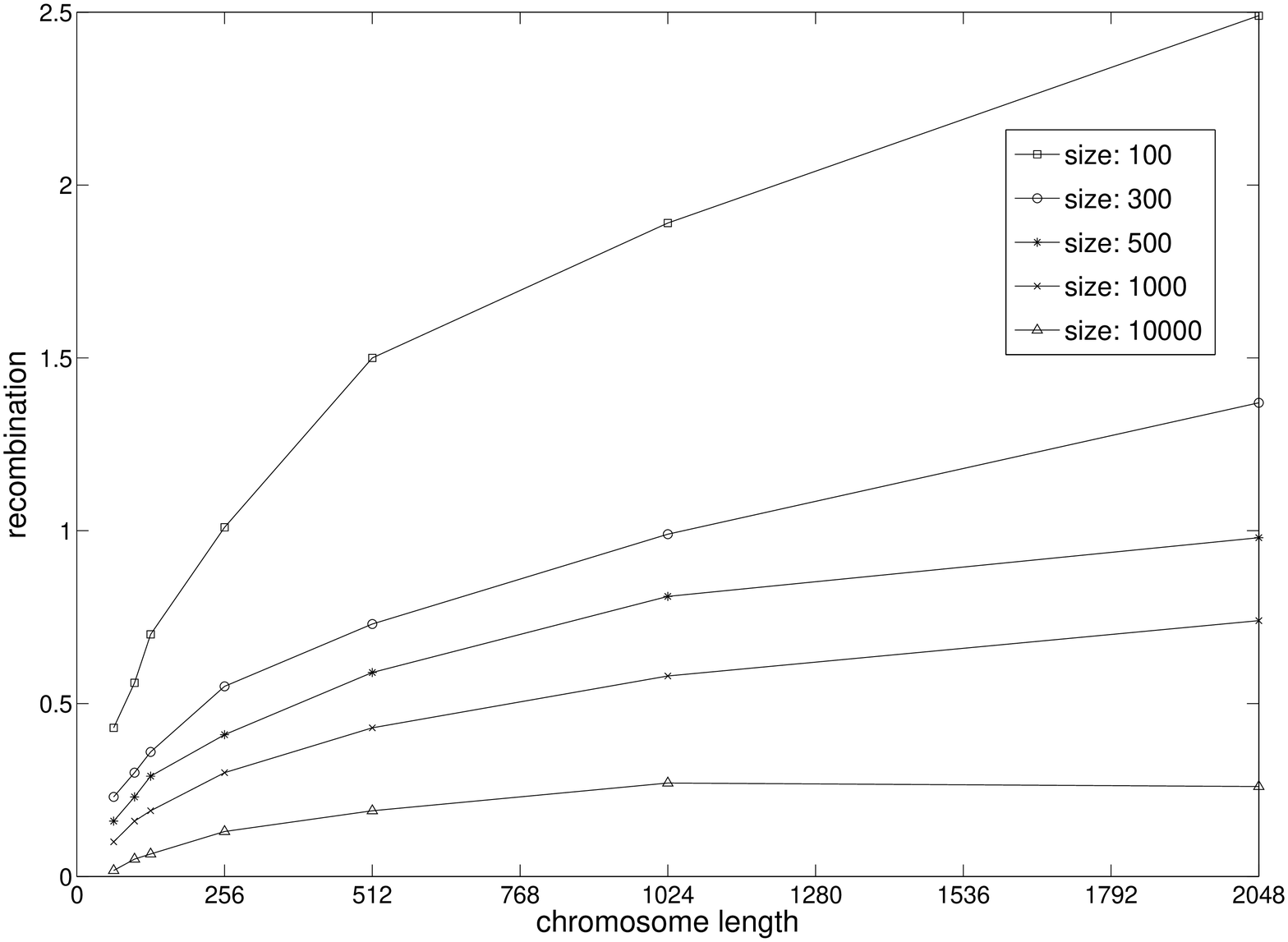,width=1\textwidth}}
  \caption{The critical recombination rates for chromosomes of different lengths and different sizes of evolving populations. Above the lines chromosomes are under purifying selection below the line are under complementation strategy.}
  \label{Figure:fig7}
\end{figure}

To understand what is going on during simulations under our parameters, take a look at Fig.\ref{Figure:fig3}. If we simulate the population without recombination - the genomes of individuals are composed of two fully complementing haplotypes $T1$ and $T2$, according to the description in the above section. $T1\||T1$ and $T2\||T2$ configurations are lethal while $T1\||T2$ and $T2\||T1$ are surviving. Imagine one recombination event between the $T1$ and $T2$ haplotypes. The recombination products, let's call them $T1xT2$, do not fit to $T1$ neither to $T2$. Thus, such products (gametes) cannot form surviving offspring in the population where they have emerged. One may conclude that introducing the recombination into the population which was earlier simulated without recombination should be deleterious for it. It is true. Introduction the crossover into the population which evolved until equilibrium without recombination has a deleterious effect on it. What would happen if the populations are evolving from the beginning at the intermediate crossover rate? We have performed a series of simulations with different crossover rates. All other parameters were constant and their values were like in the standard model. The results are shown in Fig.\ref{Figure:fig4}. There is a very sharp transition from complementing strategy, where the fraction of defective alleles is close to $0.5$, (or fraction of heterozygous loci is close to 1), to the purifying selection where the fraction of defective loci is much lower. The point of transition is around the recombination rate $0.14$. To understand better what is going on at this critical point we have analysed the relation between the fertility of individuals and crossover rate (Fig.\ref{Figure:fig5}). The criterion was the zygotic death - how many trials of forming the diploid genome of a zygote should be done to succeed in producing one surviving newborn. At the transition point (around recombination rate $0.14$) the number of trials is the highest which means that the fertility is the lowest - populations should avoid such conditions and ``escape'' to the regions where the fertility is higher - deeper in purifying strategy or in complementing strategy. The problem is ``how population can escape from those fatal conditions''. Intragenomic recombination rate is evolutionary established and it cannot be changed ad hoc. If we think how the complementing strategy is possible at all, we have to take into account the probability of meeting two complementary haplotypes, the most probably of the same origin, from the same ancestor, before they (haplotypes) mutated and recombined with other haplotypes. Before mathematicians will solve this problem formally and exactly, geneticists can say that it should depend on the size of populations. Let's try to look for a transition point in simulations performed with populations of different sizes. Results are shown in Fig.\ref{Figure:fig6}. Note, both axes are in a logarithmic scale which means that we have found a power law relation between the critical recombination rate and the size of population. The results are very interesting because of that power law. Another interesting biological question which arises is: does the value of recombination frequency at the transition point depend on the bitstring length? Fig.\ref{Figure:fig7} shows the relations between the transition points, bitstring lengths and populations' sizes. For bitstrings composed of 2048 bits, the critical recombination rate is around two crossover events per bitstring for effective population size in between 100 - 300.  The largest bitstrings in our simulations correspond to the length of the largest human chromosomes containing about 2000 genes and in fact there are two crossovers during the meiosis (gamete production) between these chromosomes, on average. Does the Nature operate close to the phase transition? Nevertheless, physicists can't say that it is a phase transition because the transition point in our case depends on the population size. But there is a trap - in all our simulations we have modelled the panmictic populations. In panmictic population an individual looks for partner for reproduction in the whole population. It is not realistic in Nature. In natural populations partners are found in close vicinity, rather. The distance or the size of groups of individuals where the individuals find their sexual partners should be considered as a real effective population size. This effective size of population determines other important genetic parameter characterizing the population - the inbreeding coefficient which is a measure of genetic relations between members of populations and more precisely between the sexual partners. 

\section{Simulations on a square lattice}
\begin{figure}
\centerline{\psfig{file=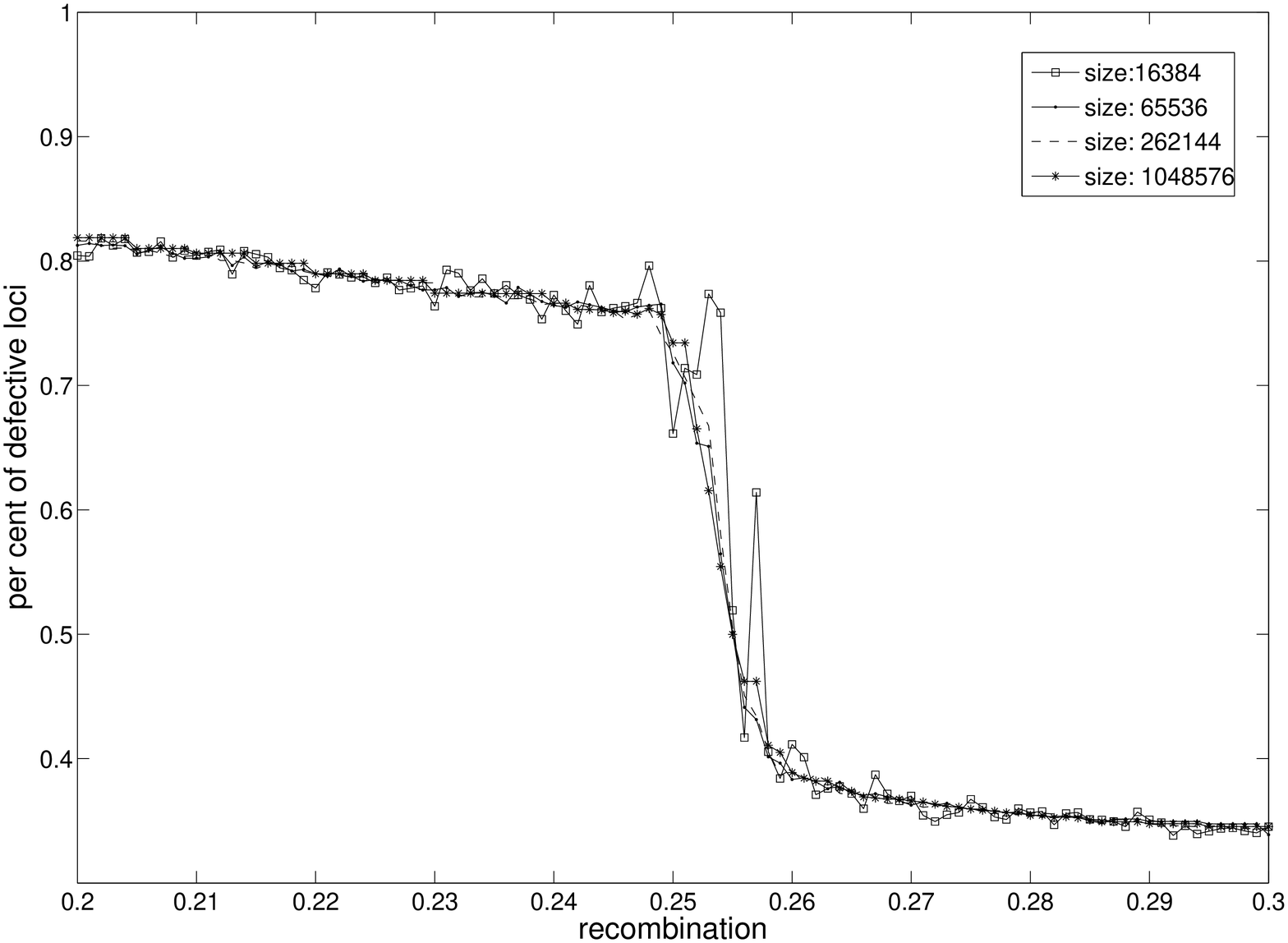,width=1\textwidth}}
  \caption{Phase transition between two strategies of the genomes' evolution - purification selection above crossover frequency 0.260 and haplotypes complementation below 0.250. Simulations were performed on square lattices (128x128 - 1024x1024) wrapped on toruses. 
The other parameters: P=B=2, L=64. Notice that transition point does not depend on the population size.}
  \label{Figure:fig8}
\end{figure}

To analyze the effect of inbreeding and effective population size on the switching between the two different strategies of evolution we have modified the model. Simulations were performed on a square lattice. Each individual occupies one field. Females are looking for their partners at a distance not larger than P and newborns are put in a free square found at a distance not larger than B from their mother. If there are no partner in the range of P and no free place at the range of B - an offspring is not born. Both distances - P and B - determine the effective size of the population (size of subpopulation where individual can look for reproduction partner), though there are some virtual borders limiting the crossbreeding but subpopulations overlaps and in fact there are no physical or geographical borders. To avoid the edge effects at the edges of the lattice (the inbreeding conditions are different because individuals have neighbours only at one side) we have wrapped the lattice on a torus. All other parameters of modelling were the same as in the previous simulations of panmictic populations. In Fig.\ref{Figure:fig8} we have presented the results of simulations for four different sizes of populations (lattices: 128x128, 256x256, 512x512 and 1024x1024). For the fixed P = B = 2 parameters, the value of critical recombination frequency does not depend on the population size - plots for all populations overlap. Note that the range of recombination frequency inside which the population changes dramatically their strategy of evolution is very narrow - between 0.250 and 0.259. The results suggest that the effect can be classified as a phase transition because the value of recombination rate at the transition point does not depend on the population size. Populations choose the more advantageous strategy on the basis of interplay between the intragenomic recombination rate and inbreeding coefficient.
The intragenomic recombination rate is an intrinsic, biological property of a genome. It is a result of a long evolution, rather and it can not be changed on a spot. On the other hand, inbreeding coefficient (the probability that the two alleles for any gene are identical by descent) could be considered as an environmental condition which easily can be changed by overcrowding or catastrophes. Thus, if populations are close to the transition point some interesting phenomena could be observed as the results of influence of the environmental conditions. For example, imagine that large panmictic population evolving under purifying selection suddenly shrinks, becomes less dense. Then, inbreeding in it could increase to such a degree that complementing strategy becomes more advantageous and, as a result, we should observe speciation.
To observe such effects we have to further modify our model on the lattice in order to recognise different species or genotypes. To accomplish that we coloured the square occupied by an individual according to the distribution of the defective alleles on its haplotypes.

Since our computer system can choose one of $2^{24}$ colours, we ascribed one colour to each of $1-2^{24}$ numbers then we cut off the central 24 bits long substring of each haplotype of a given individual (i.e. bits 21 to 44 in case of simulating the populations with L=64), converted substrings for the numbers, chosen the higher one (note that the two numbers describing two haplotypes of one genome are always different), coloured the square occupied by this very individual according to that number. The black and white edition of the paper can not show the full strength of the method - take a look at our web page, please: http://www.smorfland.uni.wroc.pl/sympatry/ 

\subsection{Sympatric speciation}
As it has been mentioned in the Introduction, sympatric speciation is an emergence of the new species inside an old one without any barriers - physical or biological. This phenomenon is still debatable because it is difficult to imagine the evolution of genetic systems which could lead to such a speciation. There are many ways to show the sympatric speciation in our model. Let's try the simplest one and start from one pair - Adam and Eve - two parents with perfect genomes (all bits set to 0) in the middle of lattice 1024x1024. The whole simulation is performed according to the standard model, with P and B distances declared. Population starts to expand on the lattice. The most critical values of simulations are P and B parameters - determining the inbreeding (environmental conditions) and, crossover rate determining the intragenomic conditions. When the P and B are low which correspond to small panmictic populations and high inbreeding, even for relatively high crossover rates the populations choose the complementing strategy (for P=B=2 critical recombination equals $0.25$, see the above section). When simulation starts from the centre of the lattice, after a short time, when genomes accumulate enough mutations to choose the strategy, the whole population is much larger than the effective populations and in fact individuals at the edges of expanding population have no chance to interbreed freely with all individuals, particularly with those at the opposite site of the territory. They evolve almost independently. Thus, the distribution of defective genes along their haplotypes is also different and they cannot form surviving offspring - they are just different species. You can observe the radiating territories of different colours. The same colours correspond to the same haplotypes' configurations, which means that individuals belong to the same species. See: http://www.smorfland.uni.wroc.pl/sympatry/. There is another possibility of observing the speciation in this system, much more beautiful and more convincing. Imagine very large panmictic population on lattice. It could evolve under relatively high P and B parameters and low recombination rate. Suddenly, the P and B parameters decrease. It could happen as a result of catastrophe when both, the density of populations and the distances covered by individuals decrease. In such conditions many of focuses of speciation arise and the spectacular speciation could be observed (like during the Cambrian era). http://www.smorfland.uni.wroc.pl/sympatry/
\pagebreak
\subsection{Some snapshots of expanding populations}

\begin{figure}[!ht]
\begin{center}
  \parbox{2.1in}{\epsfig{figure=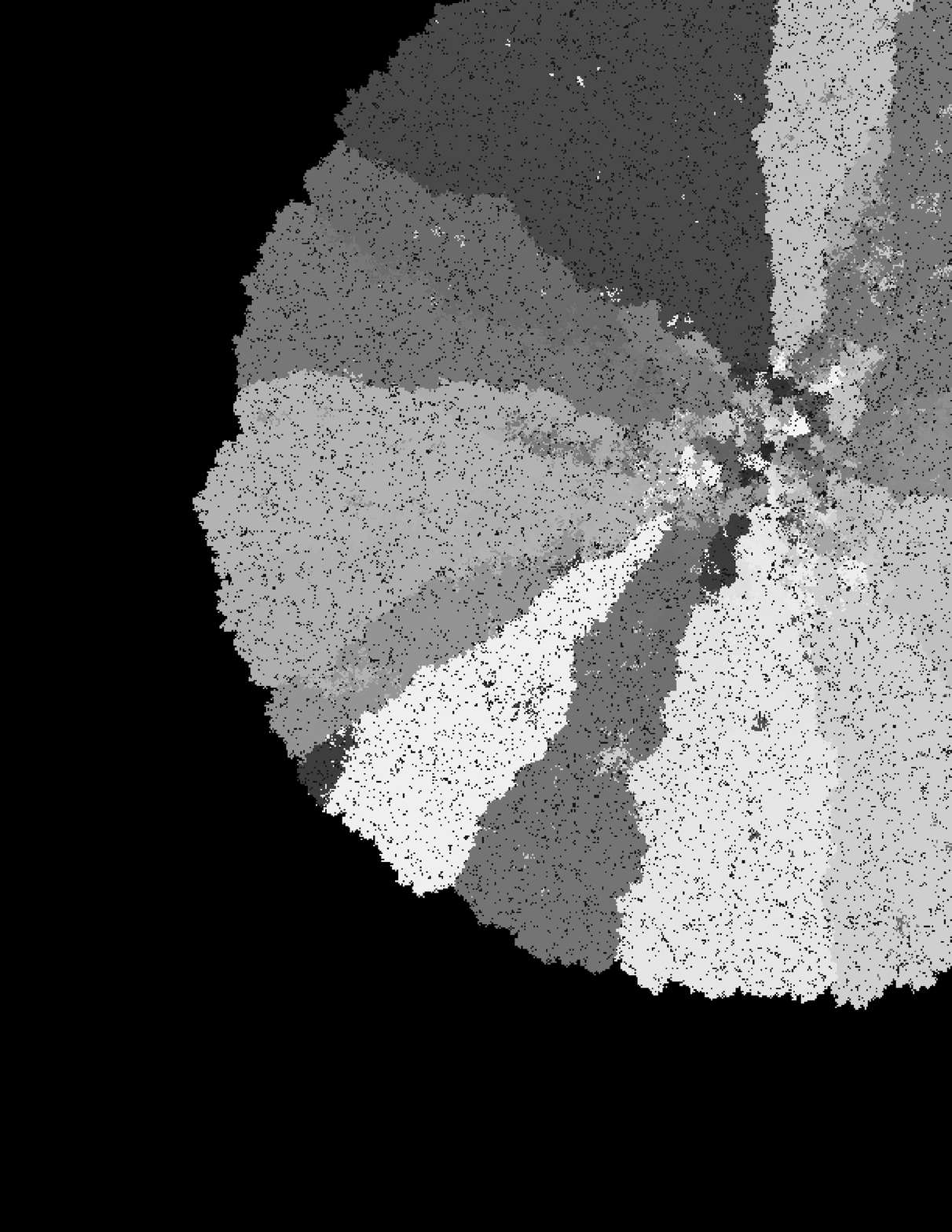,width=2in}
  \figsubcap{a}}
  \hspace*{4pt}
  \parbox{2.1in}{\epsfig{figure=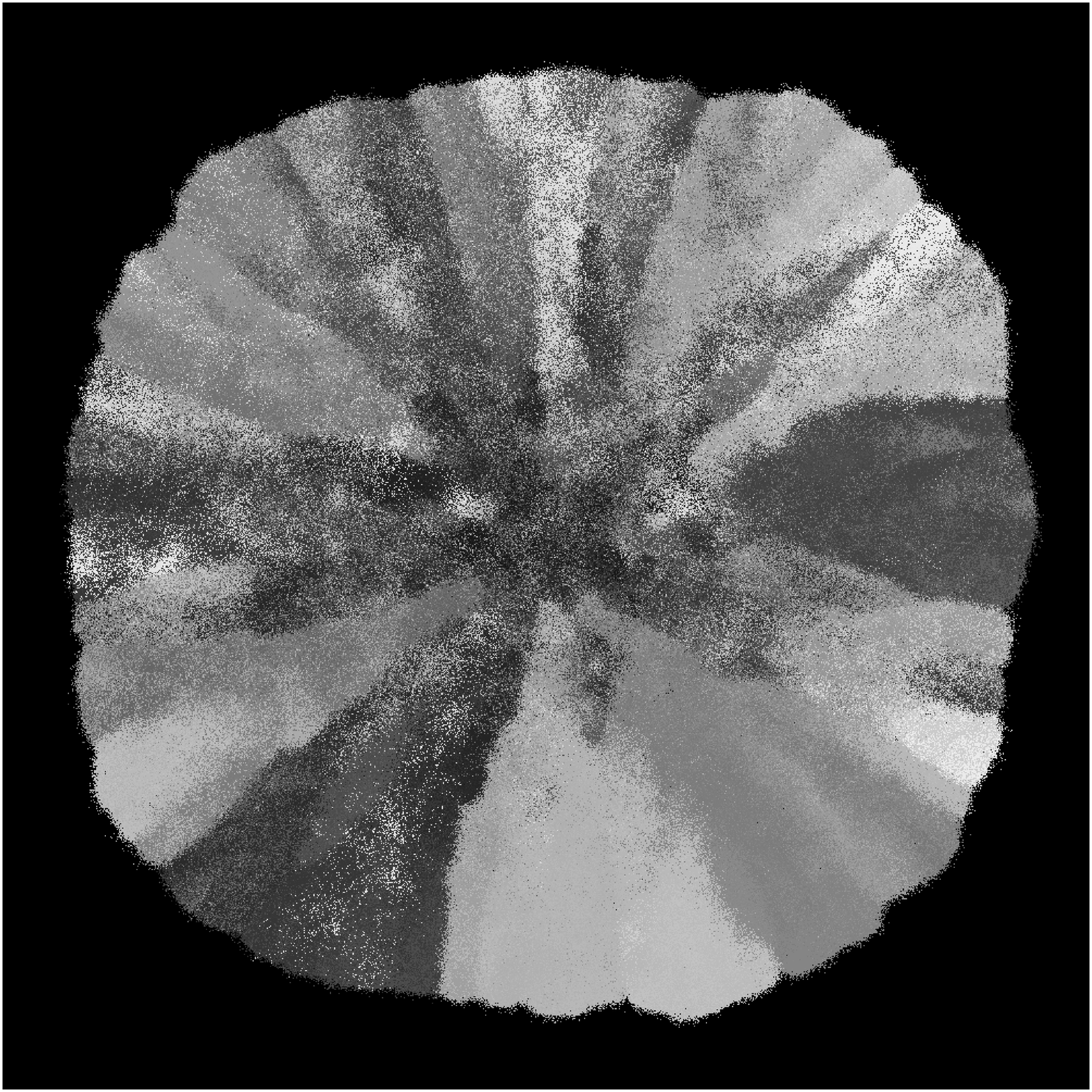,width=2in}
  \figsubcap{b}}
  \parbox{2.1in}{\epsfig{figure=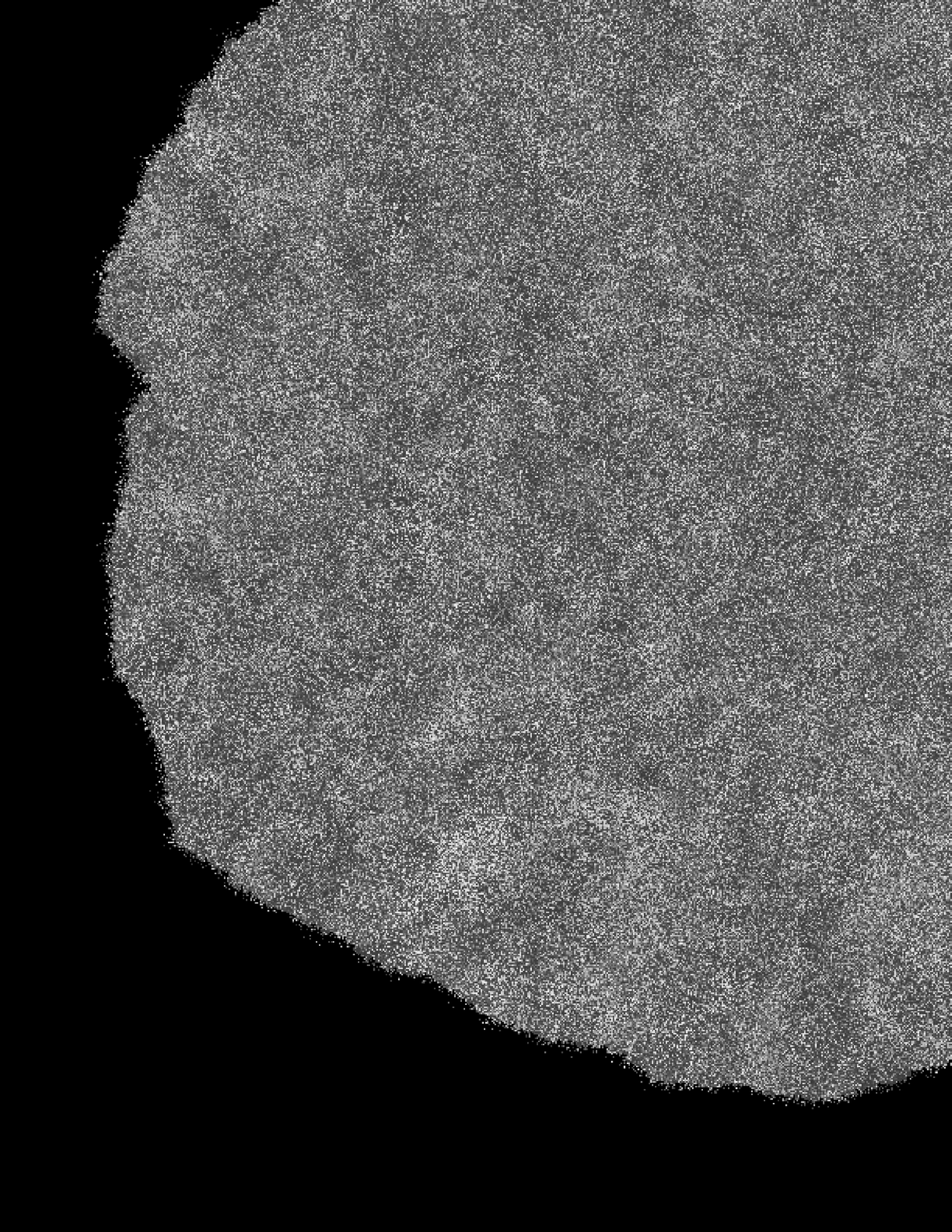,width=2in}
  \figsubcap{c}}
  \hspace*{4pt}
  \parbox{2.1in}{\epsfig{figure=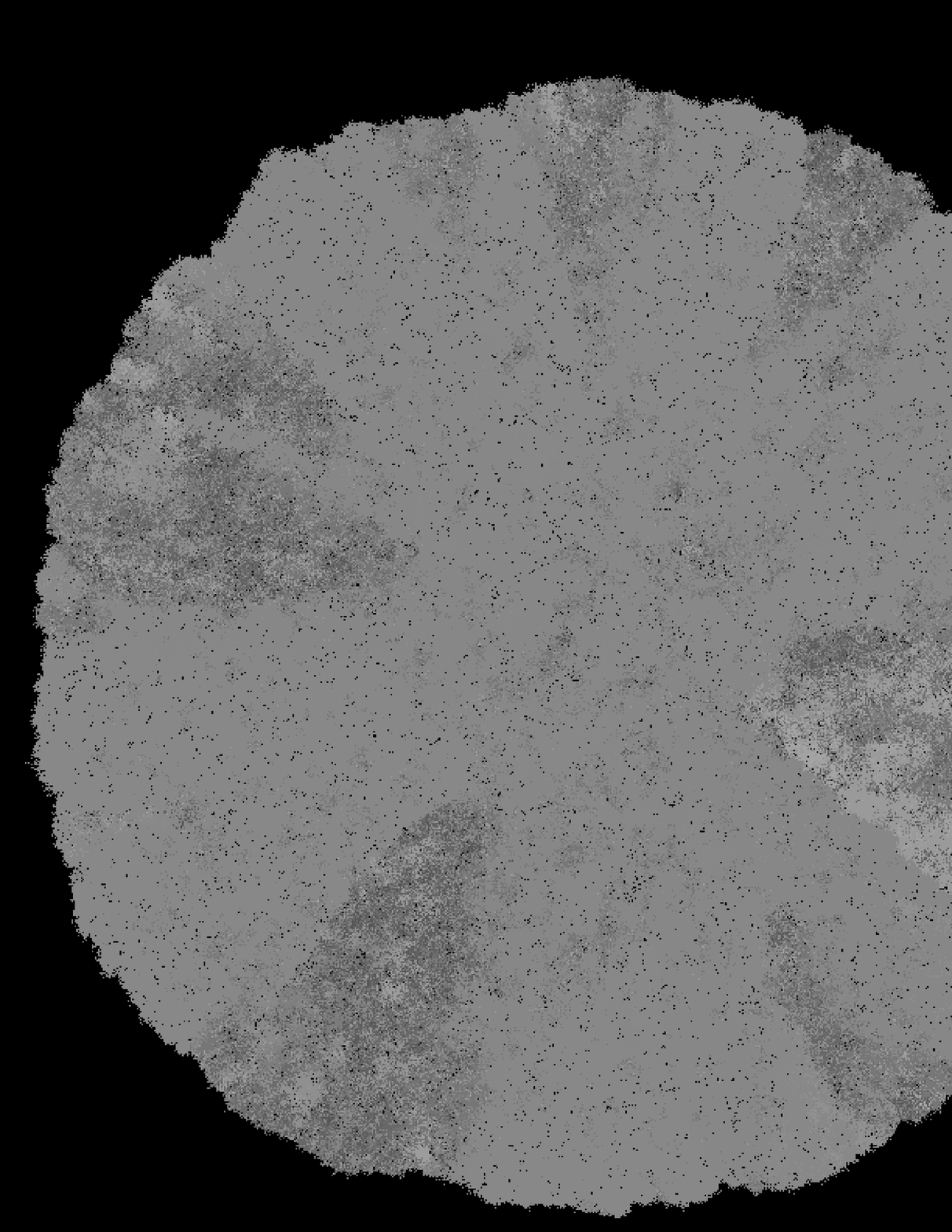,width=2in}
  \figsubcap{d}}
  \caption{Expanding populations.
  (a) Very high speciation due to a low recombination (C=0.1) and high inbred (P=B=2).
  (b) Lower inbred (P=5) and less explicit speciation.
  (c) No speciation, because of a high recombination (C=0.3) and high inbred (P=B=5)
  (d) Back-speciation from existing species.}%
  \label{Figure:populations}
\end{center}
\end{figure}

\subsection{Expansion rate and crossover frequency}
\begin{figure}
\centerline{\psfig{file=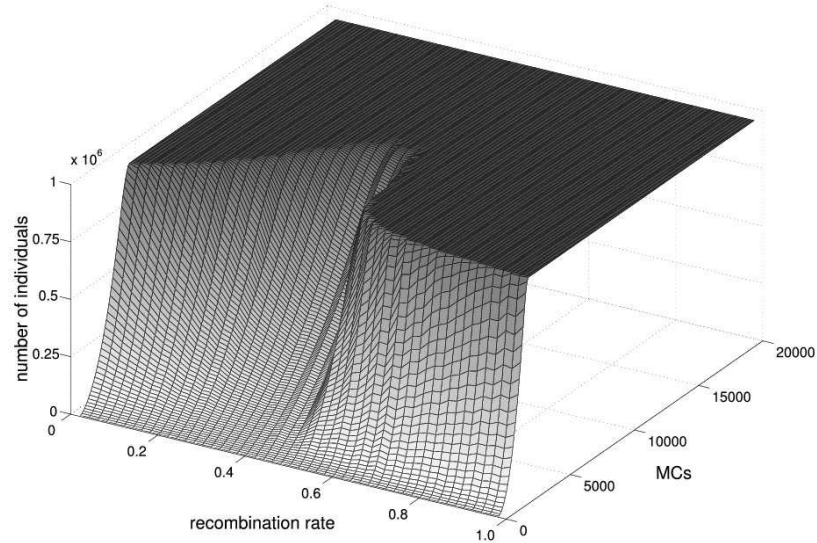,width=1\textwidth}}
  \caption{Relation between the expansion rate of population and intragenomic recombination rate.}
  \label{Figure:fig9}
\end{figure}
Let's start again with one perfect pair of parents in the middle of the lattice 1000x1000, P=B=5, crossover rate 0.1. It is obvious, that the expansion rate depends directly on P and B parameters. But at the edges it could be important how the role of the parameters deciding about inbreeding is influenced by crossover rate. One can notice that there are quite different conditions of evolution at the edges of expanding populations and inside it. At the edges, pioneers live under much higher inbreeding than inside. The question is -how the expanding rate depends on the intragenomic recombination rate under constant P and B parameters. Figure \ref{Figure:fig9}  shows the complicated, nonlinear relations between the expansion rate and crossover rate with P=B=2. The minimum of expansion is observed for crossover about 0.4. Lower or higher recombination rate increases the expansion. One can conclude, that if the population is forced to evolve longer under such conditions the recombination rate should self-organize to adapt to these conditions. It was observed in other simulations with the Penna model \cite{Cebrat}. 

\subsection{Geographical distribution of defective genes}

\begin{figure}
\centerline{\psfig{file=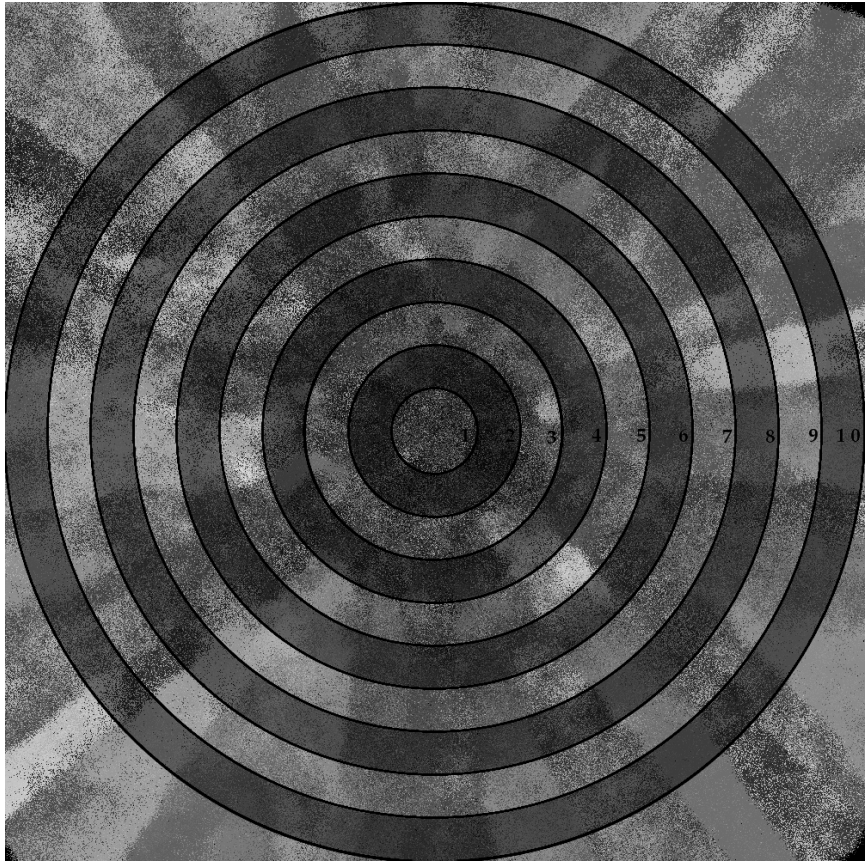,width=1\textwidth}}
  \caption{The expanding population. Rings delimit ``geographic regions'' further analysed for the defective gene distribution and the distribution of accepted crossover events (next two figures). Parameters of simulations: P=B=5, crossover rate C=0.1, L=64.}
  \label{Figure:fig10}
\end{figure}

\begin{figure}
\centerline{\psfig{file=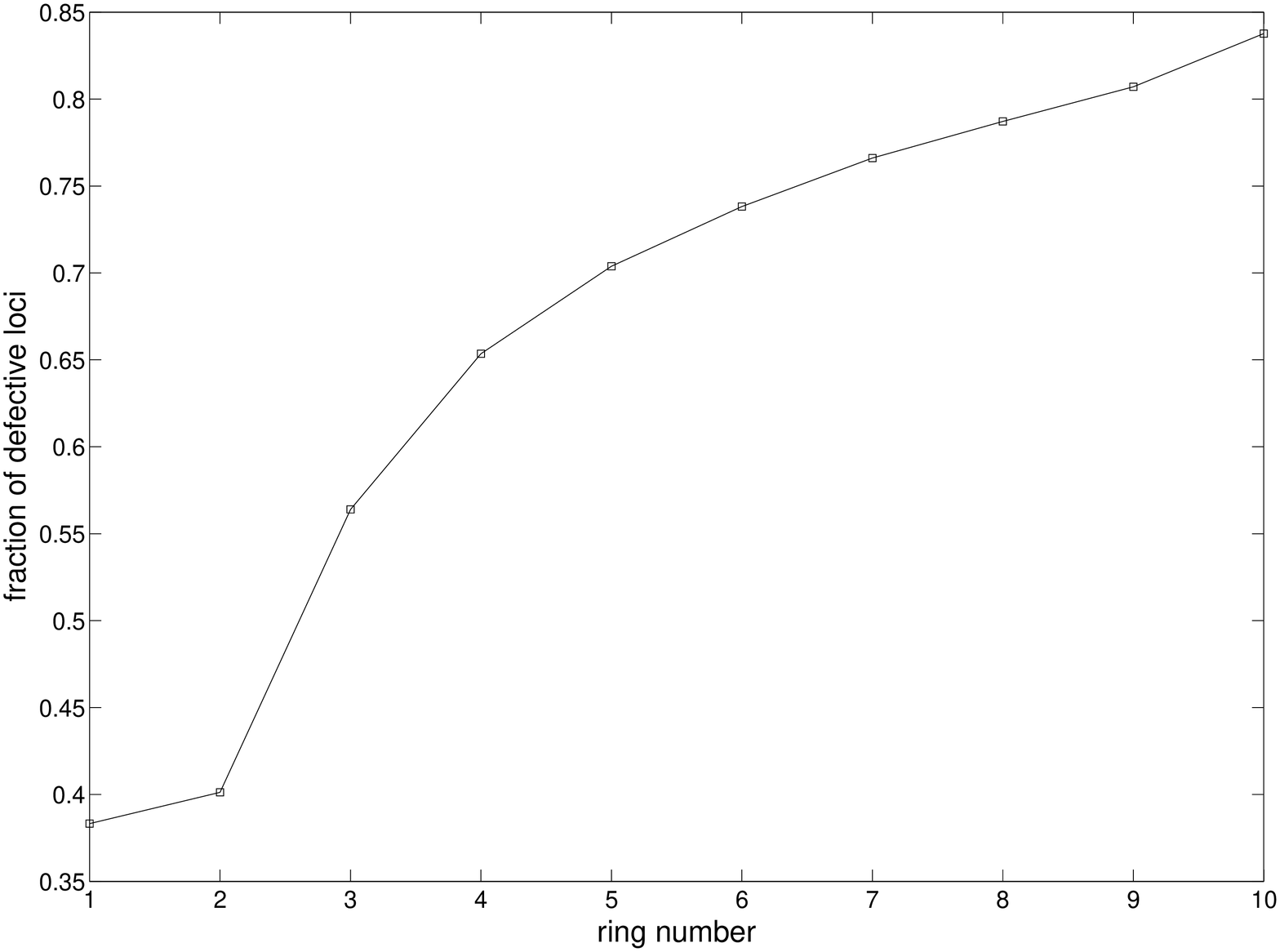,width=1\textwidth}}
  \caption{Distribution of defective genes (number of heterozygous loci) in the expanding population. The fraction of defective genes is lower in the centre of population, where for long time the whole lattice has been occupied and the inbreeding is lower. At the edges of expanding population inbreeding is higher and populations choose the complementing strategy.}
  \label{Figure:fig11}
\end{figure}

The fractions at the edges of expanding populations evolve under different condition than inside. But they are not isolated and they should influence the interior. Moreover, in our model the old individuals don't move, they stay at the same place for whole their lives, only newborns can be put aside, in the new non occupied territories. But the newborns can be put also inside if there is a place. Those complicated relations should affect the evolutionary mechanisms operating inside the population and should be seen in the genetic pool of populations. We have analyzed the population evolving under parameters: lattice size 1024x1024, L=64, M=1, P=B=5, crossover rate 0.1. Figure \ref{Figure:fig10} shows how the population looks in the early stages of evolution, still in an expansion. Figure \ref{Figure:fig11} shows the ``geographical distribution'' of defective genes in the population. At the edges of population inbreeding coefficient is higher and the complementing strategy can prevail, that is why higher fraction of defects is found in the outer parts of population. On the other hand, in the centre of expanding populations individuals reproduce under lower inbreeding for longer time thus, the purifying strategy prevails in those parts of population and the lower fraction of defects is found in this region.

\subsection{Distribution of accepted crossover events}

\begin{figure}
\centerline{\psfig{file=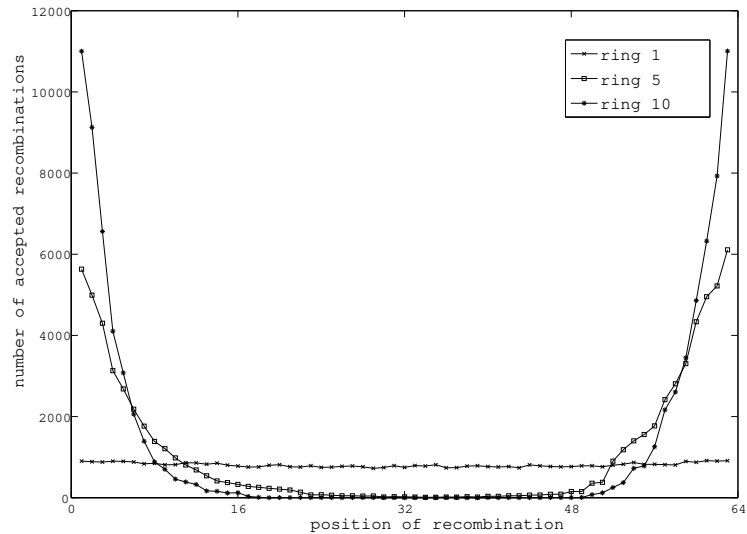,width=1\textwidth}}
  \caption{Distribution of accepted recombination events along the bitstrings in different regions of population. The probability of recombination was independent on position thus, one could expect recombination events are distributed evenly. Accepted recombination means here that haplotype after recombination formed a surviving zygote. In the centre of the population the recombination events are distributed evenly - as expected. At the regions more distant from the centre the recombinations are accepted more frequently at the ends of bitstrings which correspond sub-telomeric regions of chromosomes. }
  \label{Figure:fig12}
\end{figure}

\begin{figure}
\centerline{\psfig{file=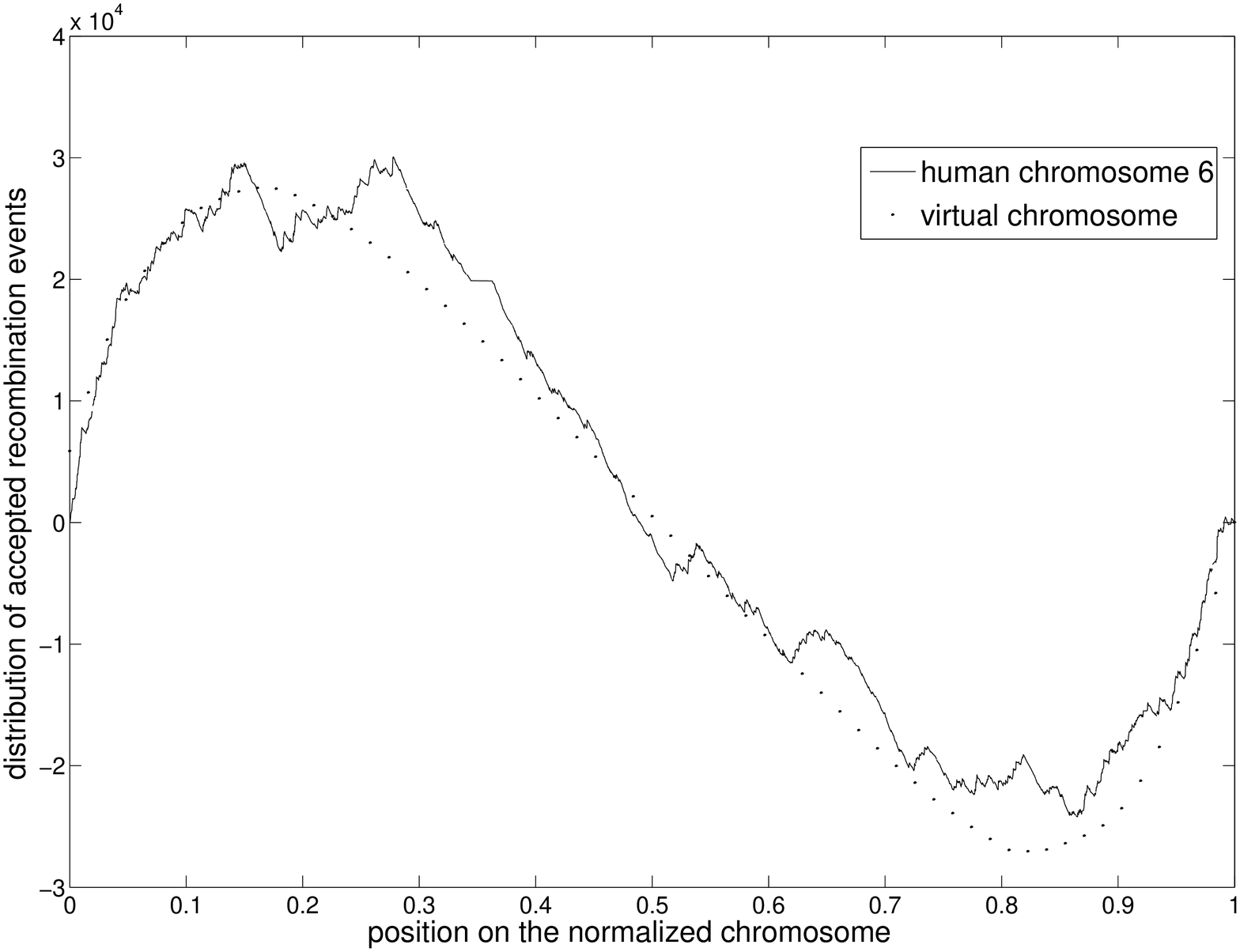,width=1\textwidth}}
  \caption{Cumulative detrended plot of frequency of accepted recombination events along the human chromosome and along the virtual chromosome. In the upward directed regions of plots the recombinations are higher than average and in the downward directed regions - lower than average.}
  \label{Figure:fig13}
\end{figure}

In the population described in the above subsection the probability of recombination between haplotypes during gamete production was 0.1. The location of recombination was randomly chosen thus, the spots of recombination events should be evenly distributed along the chromosomes in the whole pool of gametes. Nevertheless, only some gametes form the surviving zygotes. That is why we have checked if in the surviving individuals the recombinations spots are evenly distributed, as expected. We have called these recombination events - ``accepted recombinations'', since the recombination product forms the surviving individual. We have checked the acceptance of recombinations in three parts: in the centre, the fifth ring and the tenth ring. Plots in Figure \ref{Figure:fig12} show how these distributions depend on time and on the ``geographical'' location of individuals. During the first stages of evolution in the centre of population the accepted recombinations were distributed evenly. Going to the outer parts of population the characteristic bias in distribution is observed - the frequency of acceptance the recombination is higher at the terminal regions of bitstrings (in genetic terminology - in sub-telomeric regions). In the most external parts of population the recombinations in the middle of bitstrings are not accepted - it means that all recombinations which happened in this part are lost in zygotic death. How it could be explained? In the genetic language two adjacent genes are linked and the recombination probability between them is low. If we think about the bitstrings in our model it is obvious that any recombination which happens between two bitstrings separates the first bits from the last ones located at the parental bitstrings. Closer to the centre of bitstring - lower the probability of the separation of two bits by recombination somewhere between them. That is why the strategy of complementation is preserved in the centre of chromosomes longer than in the subtelomeric parts. It is a very interesting outcome of the model - the recombination events accepted by selection are not evenly distributed along the chromosomes of evolving populations. We have checked this property analysing the genomic data bases of mammals. Figure \ref{Figure:fig13} presents the detrended cumulative plots for virtual chromosome and for human chromosomes. The plots were prepared by cumulating the data of frequency of recombination in the region of chromosome diminished by the expected frequency of recombination in the region counted under assumption that recombinations are accepted evenly along the whole chromosomes or bitstrings. The distribution of accepted recombination in the natural chromosomes resembles the distribution in the virtual chromosomes - are the same mechanisms at the basis of that very important genetic property? If the answer is yes, then we can conclude that human populations (also the populations of other mammals, since the distribution of recombinations in rat's and mouse's chromosomes are of the same type) have evolved in rather small effective populations which shift the strategy of the genome evolution toward the complementing at least some parts of their genomes.

\subsection{How selection shapes the distribution of recombination spots along the chromosomes?}
\begin{figure}
\centerline{\psfig{file=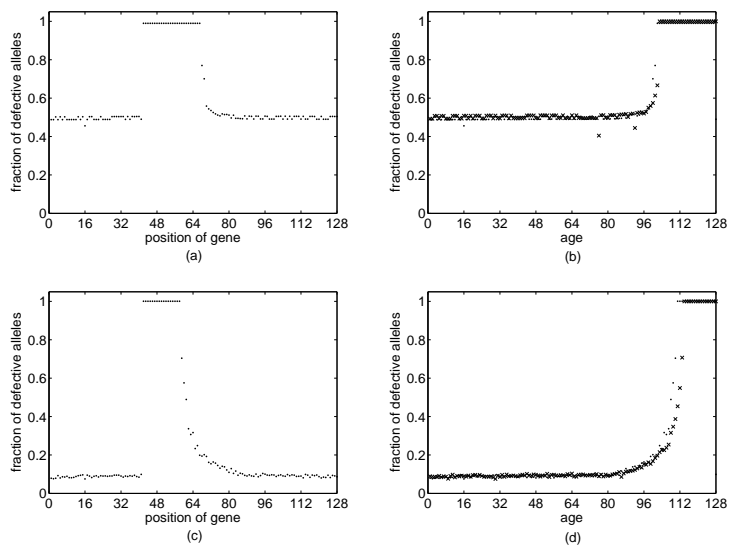,width=1\textwidth}}
  \caption{Distribution of defective genes in the genetic pool of populations evolving with recombination frequency 1 per gamete production (low panel) and, without recombination (top panel). Left panel (a,b) - the regions of haplotypes from 42 -128 bits are inverted and switched on as indicated by numbers on x-axis; a,c - recombination C=0 , b,d - recombination C=1. Right panel (c,d), the data from the left panel compared with the standard structure of genomes (non-inverted). Now x-axis is scaled accordingly to the chronology of switching on; it is in agreement with the structure for standard model only.}
  \label{Figure:fig14}
\end{figure}

\begin{figure}
\centerline{\psfig{file=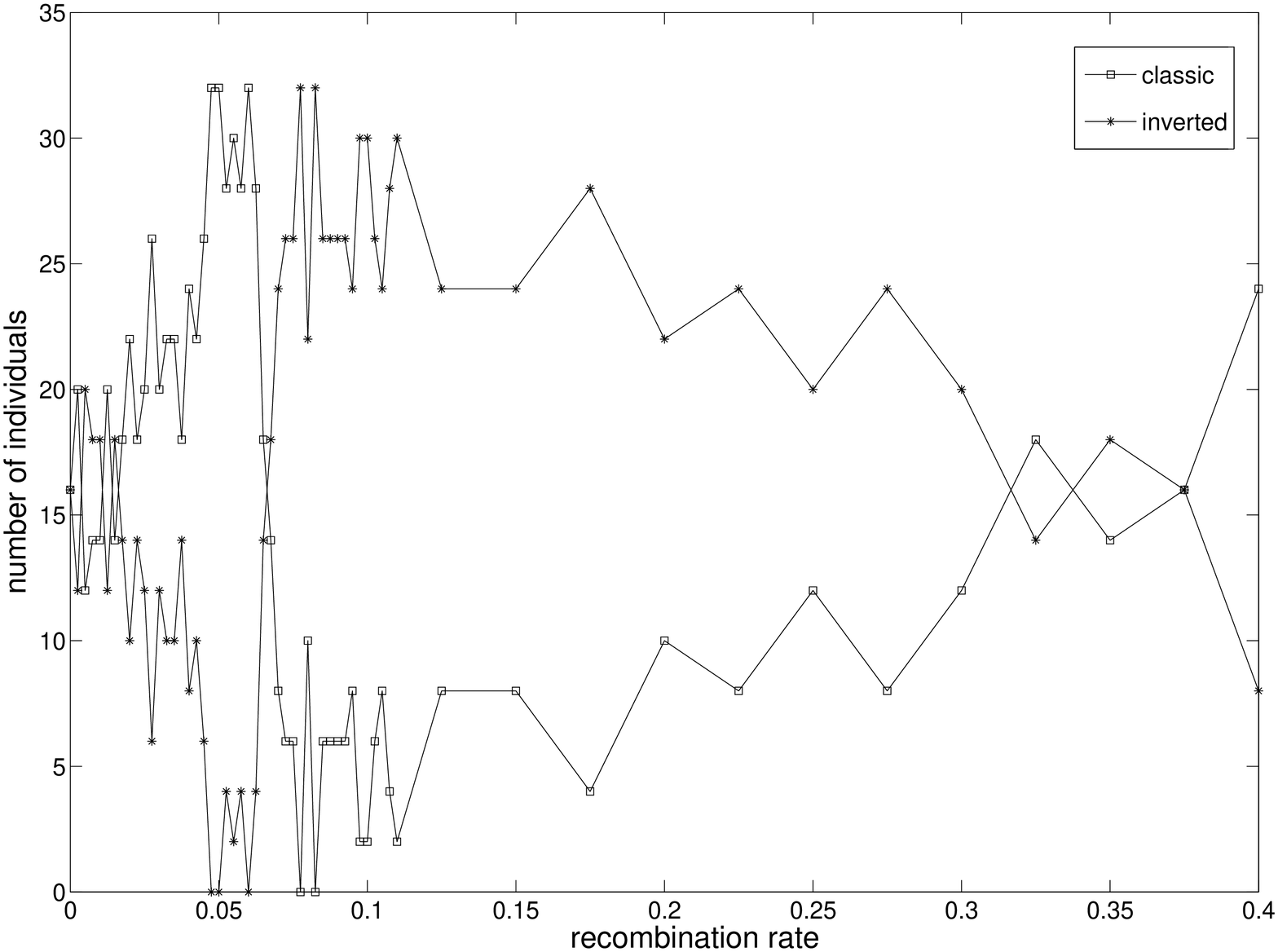,width=1\textwidth}}
  \caption{Number of winning competitions between populations with standard genome structure and populations with ``inverted'' regions of genomes. }
  \label{Figure:fig15}
\end{figure}

\begin{figure}
\centerline{\psfig{file=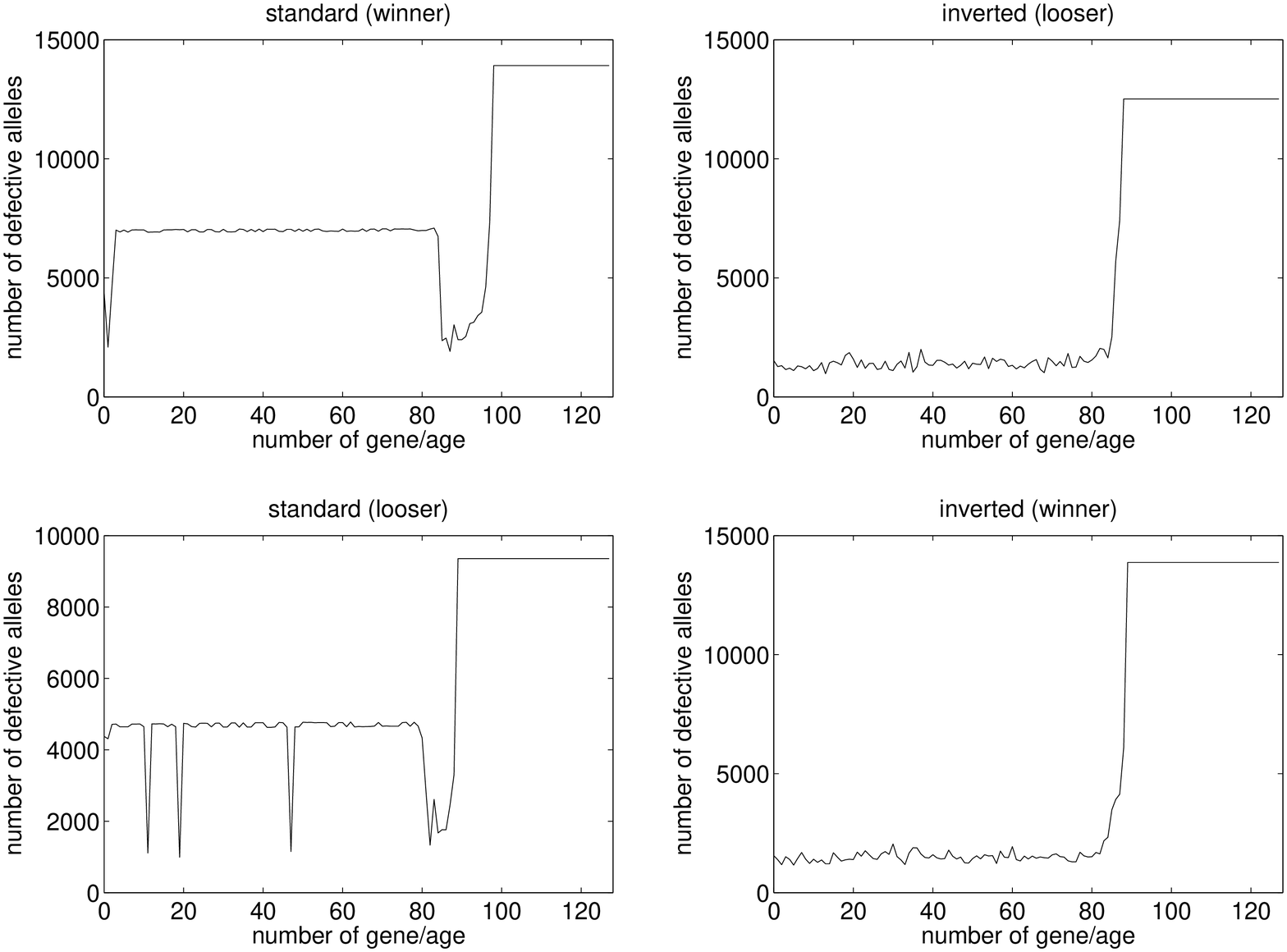,width=1\textwidth}}
  \caption{The genetic structure of pairs of winning/loosing populations. The first pair (upper) evolved under recombination rate 0.06, the second pair under recombination rate 0.08.}
  \label{Figure:fig16}
\end{figure}
In the above section we have described one of the mechanisms which could be responsible for somehow global distribution of recombination spots. Recombination spots are not evenly distributed along the natural chromosomes. There are some regions with very high recombination frequency (called hot-spots) and some regions with very low frequency of recombination (called recombination deserts) \cite{Yu}, \cite{Daly}, \cite{Jeffreys}, \cite{Arnheim}.
Is it possible that selection would favour one configuration of genome over the other one only on the basis of distribution of frequency of recombination? To check such possibility we have performed some experiments with the Penna model. Reader has to refresh his knowledge of the Chapter 2. If the simulations of population evolution are performed under parameters N=10000, L=128, M=1, B=2, R=80 and crossover rate 1 per gamete production the distribution of defective genes is characteristic for the Penna model - low and constant frequency of defects in genes expressed before the minimum reproduction age R and fraction of defects growing with age after the minimum reproduction age. All genes in the genetic pool located at position above 110 are already defective (Fig.\ref{Figure:fig14}). Thus, the last 20 positions have no effect on surviving of the individuals - at the age when those genes should be switched on, the individuals are already dead. The space occupied by these genes, from the point of view of genetic information is empty. But there is still the ``space'', it corresponds to the region of chromosome where there are no genes but where recombination could happen. 
If the recombination rate is set in simulation to 0 - the evolution drives the genomes toward the complementation strategy and the fraction of defective genes expressed before the minimum reproduction age equals 0.5 (Fig.\ref{Figure:fig14}). If the recombination rate is high - 1 per gamete production, the genomes are under purifying selection. Imagine the genome of 128 bits long but the genes are switched on in different order than in the standard model; the first 41 bits are switched on at the same order but the next region (loci 42 - 128) is inverted, thus the locus 42 is positioned now at the other end of chromosome. Is there any significant difference in the evolutionary values of such structure of genomes when compared with the standard model? Comparing the results of simulations it shows that there is no difference between the two configurations of chromosomes (Fig.\ref{Figure:fig14}). 
In simulations shown in Fig.\ref{Figure:fig14} the evolution conditions (recombination frequencies) were far from critical conditions. The critical value of recombination for these parameters of simulations is around 0.076. What could happen at critical conditions? Under such recombination rate some genes at the both ends of coding regions of chromosomes are already under purifying selection. There is a ``biological'' experiment which could allow to estimate the evolutionary values of such populations.  We can put the two populations into one environment and look for the winner. If populations have the same ``evolutionary'' value, the number of winning and loosing competitions for each population should be statistically the same. We have simulated 32 populations with a standard configuration of loci and 32 populations with the ``inverted'' second part of the genome and performed competitions of 32 pairs of such populations. The series of such 32 competitions were performed for populations evolved under different crossover frequency. The results are shown in Fig.\ref{Figure:fig15}. For recombination rate close to 0 it was no difference between the two configurations, but already for recombination $0.02$ the standard configurations won more often than the inverted one. For recombination rate in the range between $0.045 - 0.065$ the standard populations won the most of competitions ($118/128$), inverted populations were losers. Situation changed dramatically between $0.06$ and $0.075$ - in the range of recombination frequency $0.0775$ and $0.10$ the most of populations with standard configurations lost the competitions ($23/160$). Even more interesting is the distribution of defective genes in the genetic pool of winners and losers, shown in Fig.\ref{Figure:fig16}. At recombination rate of 0.06, the winning population had more defective genes expressed before the minimum reproduction age than the loosing one. But the winners complemented their haplotypes. Losers had less defective genes but they were already under the purifying selection, it was evidently more costly strategy than that of winners. Increasing the recombination rate up to $0.08$ changes the situation substantially. Now the population under purifying selection is better than population using the complementary strategy. The last one is already trying to change the strategy and larger fractions of genomes are already under purifying strategy.  Analyzing the results one has to notice that populations with inverted genome enter the purifying strategy earlier than the populations with standard genomes. Why? Because in their genomes, the defective genes which do not play any informational role (empty space) are located in the middle of the coding sequences. Any recombination inside this ``empty space'' reshuffles mutually the genes located at both ends of chromosomes and disturbs the possibility of complementation. On the other hand, in the non-inverted configuration, all recombinations which happened inside this genetically ``empty space'' had no effect on the complementation because they did not change the distribution of genes in the really coding parts of the genomes. One can expect that in natural chromosomes the space between the coding sequences can play a very important role in modelling the recombination landscape of chromosomes. 

\section{Phase transition and the length of chromosomes}
In the subsection dealing with phase transition between purifying selection and complementation strategy we have presented data concerning one pair of chromosomes. We have used bitstrings of $64$ or $100$ bits long which corresponds to chromosomes coding the same number of genes. Critical frequency of recombination was of the order of $0.1$ crossovers per pair of bitstrings which corresponds to $10$ centi Morgans (cM) in genetic units ($1$ cM corresponds to probability of crossover $0.01$). Considering the coding density as a number of genes per 1 cM  we can estimate that in our simulations it was of the order $10$ bits per $1$ cM. In human genomes number of genes per $1$ cM for different chromosomes is in the range between $2.5$ and $13.6$. It is a good agreement with our results of simulations. Generally the number of bits per $1$ cM is growing with the length of chromosome but the slope depends on the size of population. If we assume that human chromosomes evolved under conditions described by our simulation then their coding density could also corresponds to our virtual chromosomes. We have put the data corresponding to human chromosomes (coding density in relation to the size of chromosomes in genes' number). All points representing the data are found for low effective populations. Does it mean that human population have evolved under high inbreeding?

\section{Kinship and fecundity in the human population}

Fecundity (called also the total fertility rate) is a specific, very accurate measure of fertility of a given population. It corresponds to the average number of offspring produced by one female. For human contemporary populations (European) it should be around $2.08$ to fill up the gaps in our populations caused by natural and random deaths. Recently, the data concerning the fecundity of Icelander population were published \cite{Helgason}. The data covers the period from 1850 up to nowadays. Authors discovered that there is a positive correlation between the fecundity and genetic relation between spouses. The highest fecundity was observed for the genetically high related spouses. The real evolutionary success measured by the number of grand children was found for spouses related at the third - fourth cousins' level. Authors have claimed that they excluded any social and economical effects and only biological mechanisms should be responsible for their finding. 
Let's try to analyse the phenomenon from the point of view of our model. The most important assumption: the higher fecundity could be related to lower zygotic death - lower probability that a zygote does not survive until its birth or, better, until its minimum reproduction age. We have observed the phase transition in the biological reproduction - there is a value of crossover rate where, for a given size of effective population, the fecundity is the lowest (the highest number of unsuccessful trials of producing the offspring per one survival, see section and Fig.\ref{Figure:fig5}). Thus, an effect we have observed is exactly reciprocal to the expected explanation of Islander fecundity - where a maximum of fecundity was observed. Nevertheless, the transition point was characteristic for a given effective size of population whose genomes were represented by single pairs of chromosomes. In the human genomes we have 23 pairs of chromosomes of different length and coding density per recombination units. Especially the last value is an important parameter establishing the transition point. 
Let's assume the model where individuals are represented by genomes composed of two pairs of chromosomes with different coding density per recombination unit. The critical recombination value for each pair should be different, thus for a given and constant condition of recombination the critical inbreeding (the effective population size) should be different for these chromosomes.  Though, each single chromosome tends to keep the population out of its transition point and it is possible that the optimum effective size of evolving population should be somewhere in-between the critical values. If that reasoning is correct, we have to find such optimum in the model.
\begin{figure}
\centerline{\psfig{file=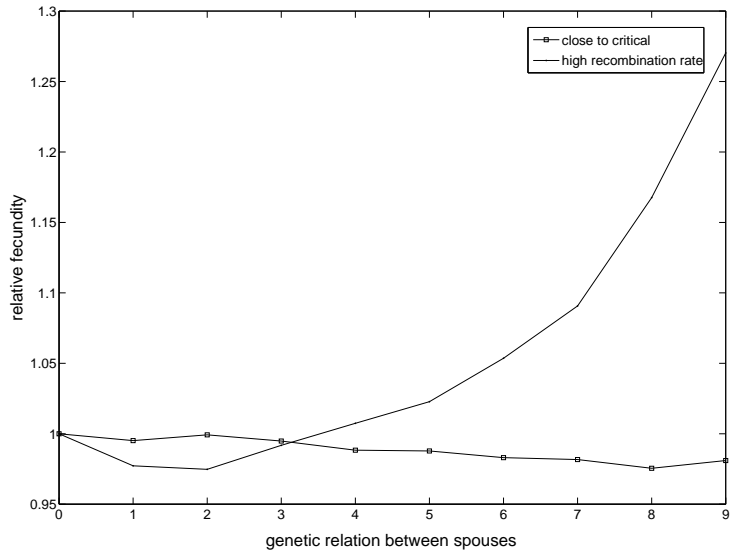,width=1\textwidth}}
  \caption{The relationship between fecundity and genetic relations between parents (x-axis show the level of cousins' relations, 0 is denoted for siblings). Growing plot is for recombination rate much higher than critical transition point, decreasing plot show the fecundity for recombination rate below the critical value for longer chromosome and above the critical value for shorter chromosome.}
  \label{Figure:fig17}
\end{figure}
\begin{figure}
\centerline{\psfig{file=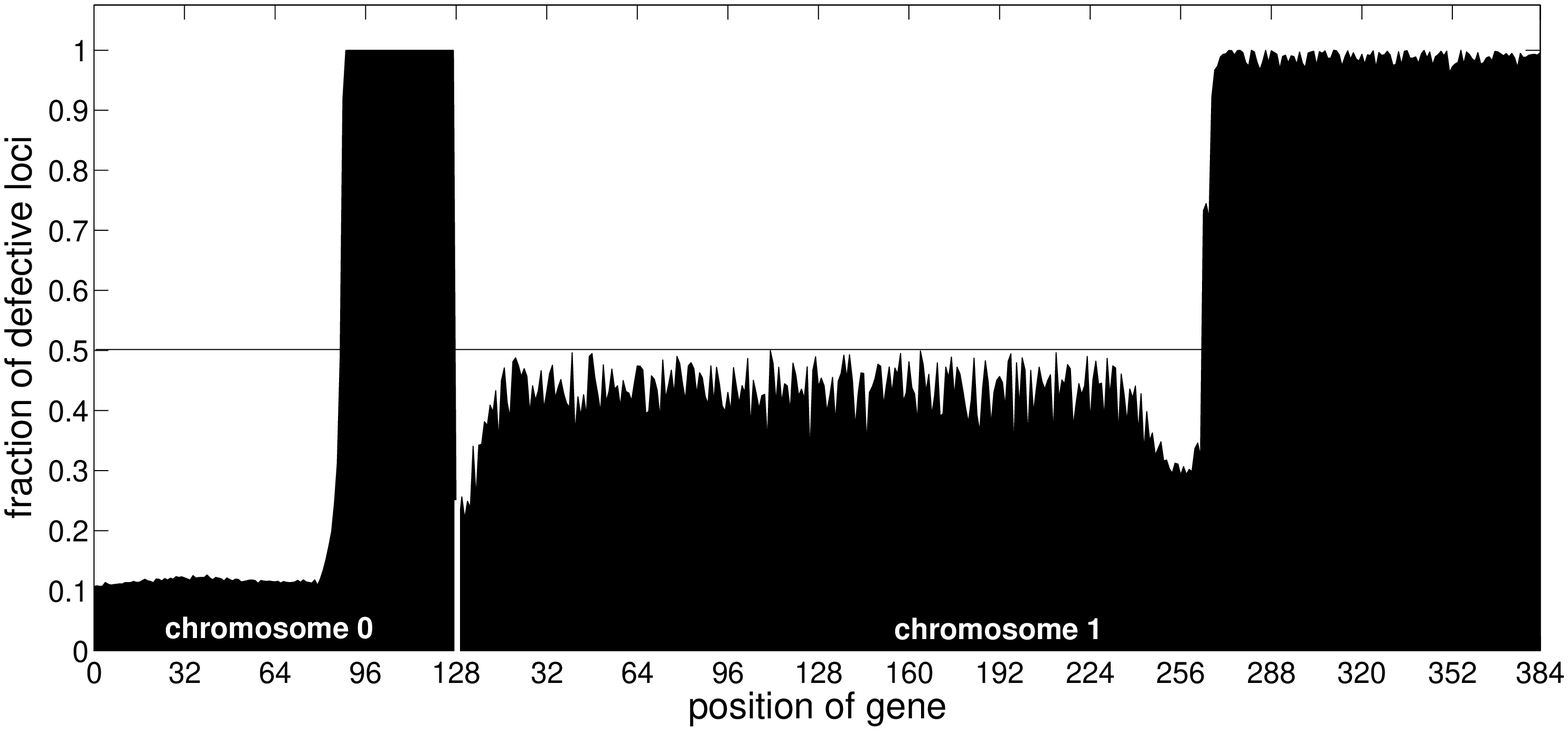,width=1\textwidth}}
  \caption{Genetic pool structure for population evolving on lattice with one chromosome 128 bits long and the second one for 384 bits long. Other parameters: recombination rate 0.24 for each pair, P=B=4. }
  \label{Figure:fig18}
\end{figure}
If the individual genomes in panmictic population (N=1000) are composed of two bitstrings L1=64 and L2=256, at the recombination frequency 0.3 the shorter chromosome is under purifying selection while the longer one under complementation strategy. In such conditions the fecundity of populations increases with the inbreeding. If the recombination is set to 1 per chromosome pair, the fecundity is anti-correlated with genetic relations between partners (Fig.\ref{Figure:fig17}).
The other preliminary data have been obtained in simulations on lattice. The parameters of the model have been chosen on the basis of data describing the relation between the critical recombination values, length of chromosomes and effective population sizes (Fig.\ref{Figure:fig5}). We have chosen chromosome pairs L1=128 and L2=384, the square lattice size 512x512 and recombination rate 0.24, P=B=4. Under such conditions the shorter chromosome stays under purifying strategy while the longer one under complementary strategy (Fig.\ref{Figure:fig18}). Further studies are connected with looking for conditions where non-monotonous relation for real evolutionary costs could be obtained. We expect that in shrinking population (higher inbreeding) selection would try to push the shorter chromosome under complementing strategy (bad situation - higher zygotic mortality). Increasing the effective population (lower inbreeding) would try to push the longer chromosome toward the purifying selection - also bad situation, again higher zygotic death frequency. What is the solution? To stay with the optimum of inbreeding - population can grow but the traditional reproduction habits should be kept or restored - they supposed to be the best. 

\pagebreak
 
\bibliographystyle{ws-rv-van}

\section*{Acknowledgements}
Calculations have been carried out in Wroc\l{}aw Centre for Networking and Supercomputing (http://www.wcss.wroc.pl), grant \#102.

\printindex                         
\end{document}